\documentclass[%
 reprint,
 amsmath,amssymb,
 aps,
]{revtex4-1}

\usepackage{hyperref}
\usepackage{amsmath}
\usepackage{dblfloatfix}    

\usepackage{microtype} 

\usepackage[utf8]{inputenc}
\usepackage[T1]{fontenc}
\usepackage{lmodern}
\usepackage{units}
\usepackage{setspace}
\usepackage{fixltx2e}
\usepackage{amsfonts}
\usepackage{amssymb}
\usepackage[dvipsnames]{xcolor}

\usepackage{tikz}
\usepackage{pgfplots}
\tikzset{ font=\footnotesize}
\usetikzlibrary{
    external,
    arrows,
    backgrounds,
    calc,trees,shadows,positioning,arrows,chains,fadings,
    decorations.pathreplacing,
    decorations.pathmorphing,
    decorations.shapes,
    decorations.text,
    shapes,
    shapes.geometric,
    shapes.symbols,
    matrix,
    patterns,
    intersections,
    fit,
    spy
}

\begin{document}

\preprint{APS/123-QED}

\title{Origin of reversible and irreversible atomic-scale rearrangements in a model two-dimensional network glass}

\author{Firaz Ebrahem}\email{ebrahem@iam.rwth-aachen.de}
\author{Franz Bamer}
\author{Bernd Markert}

\affiliation{Institute of General Mechanics, RWTH Aachen University, 52062 Aachen, Germany}



\date{\today}

\begin{abstract}
In this contribution, we investigate the fundamental mechanism of plasticity in a model two-dimensional network glass. The glass is generated by using a Monte Carlo bond-switching algorithm and subjected to athermal simple shear deformation, followed by subsequent unloading at selected deformation states. This enables us to investigate the topological origin of reversible and irreversible atomic-scale rearrangements. It is shown that some events that are triggered during loading recover during unloading, while some do not. Thus, two kinds of elementary plastic events are observed, which can be linked to the network topology of the model glass.
\end{abstract}


\maketitle

\section{INTRODUCTION}

It is generally accepted that plastic deformation in glasses is triggered by atomic-scale rearrangements, referred to as shear transformations \cite{Spaepen1977,Argon1979b,Falk1998}. These inelastic events are localized in space and they involve only a small number of atoms. In recent years, much effort has been devoted to achieving a deeper understanding of the atomic phenomena associated with the plastic deformation of glasses. In this regard, numerical studies have been performed extensively on two-dimensional binary glass forming systems \cite{Tomida1993,Lacks2004,Maloney2006,Tanguy2006,Tsamados2009,Patinet2016,Barbot2018}. To investigate how pure structural disorder correlates with plastic response, the athermal quasistatic (AQS) deformation protocol has proved useful. The glass system is subjected to small shear increments, each followed by energy minimisation. Thus, the AQS protocol is used to explore the potential-energy landscape (PEL) \cite{Heuer2008,Rodney2011,Hufnagel2016}. The AQS deformation  allows the system to overcome the energy barrier to jump from one local minimum of the potential energy to another neighboring minimum. According to the PEL, the atomic-scale rearrangements can be considered as instabilities of the glass system. The mechanical stability is governed by the Hessian matrix $\boldsymbol{H}$, a matrix of the second-order partial derivatives of the total potential energy of the system $U(\boldsymbol{r}_1, \boldsymbol{r}_2, \dots, \boldsymbol{r}_N)$, which is a function of the particle positions $\{ \boldsymbol{r} \}^{N}_{i}$. The elements of the Hessian matrix are given as:
\begin{eqnarray}
 H_{ij} = \frac{\partial^2 U(\boldsymbol{r}_1, \boldsymbol{r}_2, \dots, \boldsymbol{r}_N)}{\partial \boldsymbol{r}_i \partial \boldsymbol{r}_j} \, .
 \label{eq:hessian}
\end{eqnarray}
As long as the system is mechanically stable, the Hessian matrix $\boldsymbol{H}$ is symmetric and positive semidefinite. It has zero eigenvalues associated with the Goldstone modes, while all the remaining eigenvalues are real and positive \cite{Ashwin2013,Gendelman2015}. The plastic instability under applied shear strain $\gamma$ occurs via a saddle-node bifurcation, with the lowest positive eigenvalue $\lambda_{\text{min}}$ going to zero as 
\begin{equation}
\lambda_{\text{min}} \propto \sqrt{\gamma_c - \gamma} \, ,
\label{eq:lambda}
\end{equation}
where $\gamma_c$ it the critical shear strain at the bifurcation. As the lowest eigenvalue $\lambda_{\text{min}}$ of the Hessian matrix $\boldsymbol{H}$ approaches zero the corresponding eigenfunction localizes on the typical quadrupolar-like structure. This quadrupolar symmetry is similar to an Eshelby inclusion embedded in an elastic surrounding \cite{Eshelby1957,Albaret2016,Hieronymus_Schmidt2017,Cao2018}. It was shown that the nonaffine irreversible displacement field at the plastic event scales with the eigenfunction whose eigenvalue vanishes at the instability \cite{Maloney2004a,Maloney2004b,Lemaitre2007,Rodney2011}. 

The plastic events are identified from the stress-strain curves during AQS shear deformation of glass samples. The stress-strain curves are made of elastic segments intersected by sudden stress drops, corresponding to the localized rearrangements of a subset of atoms. Shear loading and unloading simulations of metallic glasses revealed that some events are recovered during the unloading process, while other events are not \cite{Xu2017,Xu2018}. From these loading unloading investigations Xu et al.~concluded that in glasses, at least two kinds of plastic events can be identified, as originally proposed by Argon and Shi \cite{Argon1983}. The local atomic rearrangements that recover upon unloading have been referred to as local ``anelastic'' transformations. The other local rearrangements that can not recover upon unloading were identified as ``viscoplastic'' transformations. However, the fundamental mechanism of these two types of elementary plastic events in glasses is, to this date, not fully resolved. In particular, for network glasses, such as vitreous silica, there is a lack of knowledge about the microscopic deformation mechanism, although these glasses can be well defined through the network topology, which is, in two dimensions, governed by corner-sharing SiO$_3$ triangles \cite{Morley2018,Morley2019,Bamer2020}.

Thus, in this contribution, we study the origin of reversible and irreversible atomic rearrangements in a model two-dimensional (2D) network glass. The model glass is numerically prepared by applying a series of topological flip transformations to an initially hexagonal lattice. The network structure is generated based on statistical data extracted from real 2D silica \cite{Lichtenstein2012a}. We investigate the response of the model glass under AQS simple shear deformation and demonstrate that the two kinds of elementary plastic events (reversible and irreversible) can directly be linked to the network topology of the glass.

\section{METHODS}

To generate the model 2D network glass, we use a recently developed potential function that has been introduced to describe silica glass in a two-dimensional framework \cite{Roy2018,Roy2019a,Roy2019b,Bamer2019b,Ebrahem2020}. Thus, the model glass used in this study meets the Zachariasen requirements of network glasses consisting of covalently bonded corner-sharing SiO$_3$ triangles \cite{Zachariasen1932}. The interatomic interactions are described by a Yukawa-type potential function
\begin{equation}
 U_{ij} = \left( \frac{\sigma_{ij}}{r_{ij}} \right)^{12} + \left( \frac{q_{ij}}{r_{ij}} \right) \, \text{exp}\left( -\kappa r_{ij}  \right) \,,
 \label{eq:yukawa}
\end{equation}
where the parameters $\sigma_{ij}$, $r_{ij}$, $q_{ij}$ and $\kappa$ denote the repulsive strength, the interatomic distance, the ionic charge and the screening parameter, respectively. The parameter values can be taken from Roy et al.~\cite{Roy2018}. This potential function was shown to successfully reproduce both the short-range order and the ring statistics of real 2D silica \cite{Morley2018,Morley2019,Bamer2020}. The potential is dimensionless. Thus, all quantities except the length scales are expressed in dimensionless units.  Furthermore, the potential is truncated at $11$ \r A and force-shifted in order to ensure that both the potential energy and the corresponding interatomic forces vanish smoothly at the cut-off distance $r_c$. This is achieved by adding a linear term to the potential \cite{Allen1989}. For the shifted potential $\bar{U}_{ij}$, it follows
\begin{equation}
 \bar{U}_{ij} = 
 U_{ij}(r_{ij}) - U_{ij}(r_c) - ( r_{ij} - r_c) \left.\dfrac{\mathrm{d}U_{ij}}{\mathrm{d}r_{ij}} \right|_{r_{ij}=r_c}
 \label{eq:force-shifted}
\end{equation}
for $r_{ij} \leq r_c$ and $\bar{U}_{ij}=0$ otherwise.

To generate the glass samples, we start from the hexagonal lattice, as observed for crystalline 2D silica \cite{Todorova2006}. The hexagonal lattice considered in this study consists of 1350 atoms (540 silicon and 810 oxygen atoms) that form 270 six-membered rings in a periodic simulation cell with a size of about $8 \times 8$ nm$^2$. The hexagonal lattice is then transformed into the underlying dual lattice that is subjected to a Monte Carlo bond-switching algorithm by applying in-plane flips of selected dual  bonds \cite{Morley2018,Bamer2020}. After each flip, the system energy is minimized using a harmonic dual potential describing ring-ring interactions \cite{Ebrahem2020b}. This operation affects a set of four neighboring rings in the network by increasing the size of two rings and decreasing the size of the other two rings. After a sequence of random flips, the dual is transformed back to the overlying network and minimized by using the Yukawa-type potential function (\ref{eq:yukawa}).

Furthermore, in order to avoid local ring clusters of similar sizes in the glass samples, we use the empirical Aboav--Weaire law \cite{Aboav1970,Aboav1980,Weaire1974} that controls the ring neighborhood statistics:
\begin{equation}
 m_n = \mu \left( 1- \alpha_n \right) + \frac{1}{n} \left( \mu^2 \alpha_n + s^2 \right) \, ,
\end{equation}
where $m_n$ denotes the mean ring size around a ring of size $n$, and $\mu$ and $s^2$ are the mean and the variance of the ring size distributions, respectively. The ring neighborhood is then identified by only one Aboav--Weaire parameter $\alpha_n$. Thus, proper values of $\alpha_n$ lead to physically meaningful ring neighborhood distributions and, therefore, prevent an unphysical clustering effect of small or large rings. The images of real 2D silica have been shown to reveal a fully coordinated structure \cite{Lichtenstein2012a}, proving Zachariasen's network theory. In two dimensions, one silicon atom of our SiO$_{1.5}$ network glass is always surrounded by three oxygen atoms, such that no dangling bonds are observed \cite{Roy2018,Roy2019a,Roy2019b}. Thus, the presented bond-switching algorithm produces perfect 2D glass models with fully coordinated networks according to Zachariasen's random network theory. The glass samples obtained by using this strategy are then subjected to AQS simple shear loading and unloading, applying Lees-Edwards boundary conditions. The deformation is applied in sufficiently small shear increments of $\Delta \gamma = 4 \times 10^{-6}$. Each shear increment is followed by energy minimisation with a tolerance of $10^{-16}$, using the conjugate gradient method.

\section{RESULTS AND DISCUSSIONS}

\begin{figure}[]
\centering
\includegraphics[width=8.6cm]{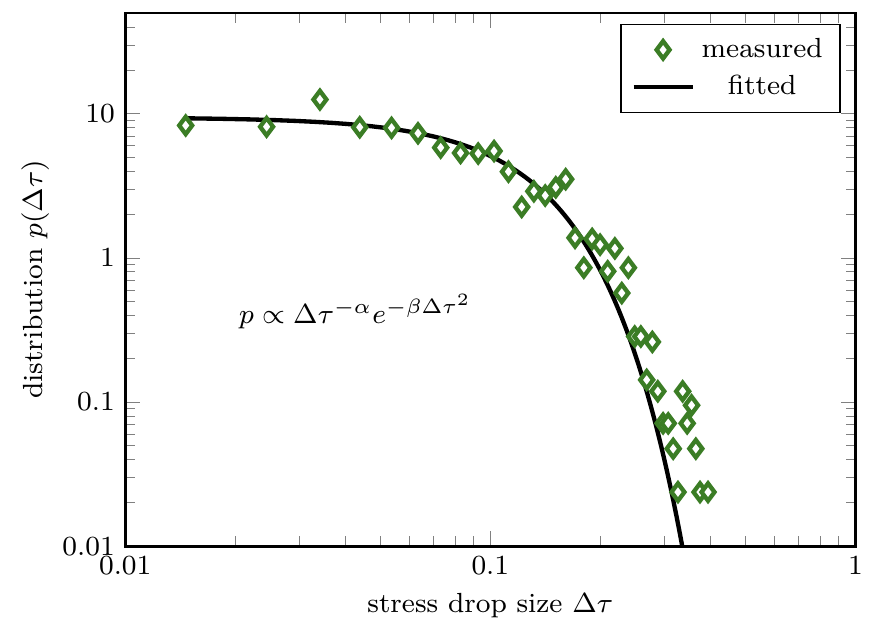}
\caption{Stress drop distribution of the model 2D network glass under simple shear loading. Logarithmic plot of the probability distribution of the stress drop size. The black solid line represents the fitted power law $p \propto \Delta \tau^{- \alpha} e^{- \beta { \Delta \tau }^2 }$ of the measured stress drop sizes (green diamond markers).}
\label{fig:aval_stat}
\end{figure}

In a first simulation set, a total number of 100 samples are subjected to simple shear loading up to a strain of $\gamma = 100 \, \%$. The shear stress drops are identified from the  stress-strain curves and their sizes are measured, while the stress is normalised with respect to the maximum occurring value of all samples. Based on this simulation set, we perform a statistical evaluation of the events. The corresponding statistics of the stress drop size is shown in Fig. \ref{fig:aval_stat}, where the green diamond markers represent the histogram of the stress drop sizes, subdivided into 40 intervals. It is shown that the distribution follows a power law of the form $p \propto \Delta \tau^{- \alpha} e^{- \beta { \Delta \tau }^2 }$, where $\alpha$ and $\beta$ are free constants. This finding is in very good agreement with previous studies on amorphous plasticity \cite{Talamali2011,Nicolas2018}.

\begin{figure}[]
\centering
\includegraphics[width=8.6cm]{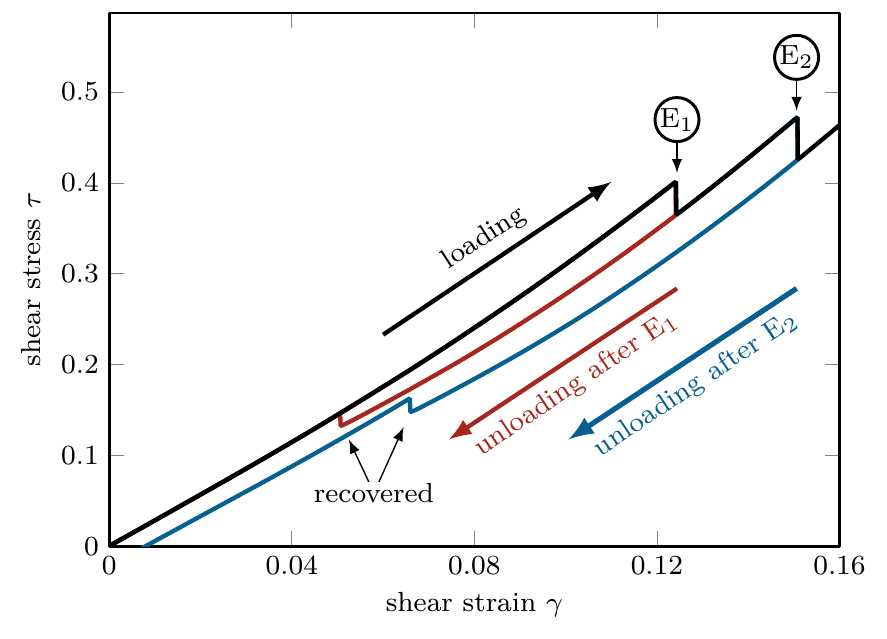}
\caption{Stress-strain relation of a model 2D network glass under AQS simple shear deformation. The black curve corresponds to the simple shear loading up to a strain of about $16 \, \%$, where two stress drops E$_1$ and E$_2$ are identified. The red curve corresponds to the stress response to shear unloading directly after the event E$_1$. The blue curve corresponds to the stress response to shear unloading directly after the event E$_2$.}
\label{fig:sig_eps_sample_1_reload}
\end{figure}

In a second simulation set, the stress response at smaller strain values is addressed in order to investigate the initial plastic events. In what follows, one representative sample will be discussed in detail, followed by an extended statistical analysis by averaging over 100 samples. Figure \ref{fig:sig_eps_sample_1_reload} shows the stress response of the representative sample under AQS simple shear deformation. The model 2D network glass is subjected to shear loading up to a strain of $\gamma = 16 \, \% $. In this way, two sudden stress drops are identified from the shear stress-strain loading curve (black curve in Fig. \ref{fig:sig_eps_sample_1_reload}), marked by the deformation states E$_1$ and E$_2$.

Each stress drop corresponds to an elementary plastic event. Once an elementary plastic event is identified in the first AQS simulation, the atomic configuration right after each stress drop is chosen as a starting point for subsequent simulations, in which the model glass is unloaded by reversing the simple shear deformation. The stress response of the first AQS unloading simulation, i.e., unloading after the first event E$_1$, is shown as the red curve in Fig. \ref{fig:sig_eps_sample_1_reload}. It can clearly be seen that the first elementary plastic event is recovered at a strain of around $5 \, \%$ during shear unloading. In the next step, shear unloading is started right after the second event E$_2$. The corresponding stress-strain relation is shown as the blue curve in Fig. \ref{fig:sig_eps_sample_1_reload}. Observation of this curve reveals that although two elementary plastic events E$_1$ and E$_2$ are triggered during shear loading, only one event is recovered during shear unloading when the latter is started right after the second event E$_2$. This observation is in line with other atomic-scale studies that reported at least two kinds of shear transformations in metallic glasses: one of them as a local ``anelastic'' transformation that can be recovered upon unloading, while the other one is a ``viscoplastic'' transformation that cannot recover upon unloading \cite{Argon1983,Xu2017,Xu2018}.

\begin{figure}[]
\centering
\includegraphics[width=8.6cm]{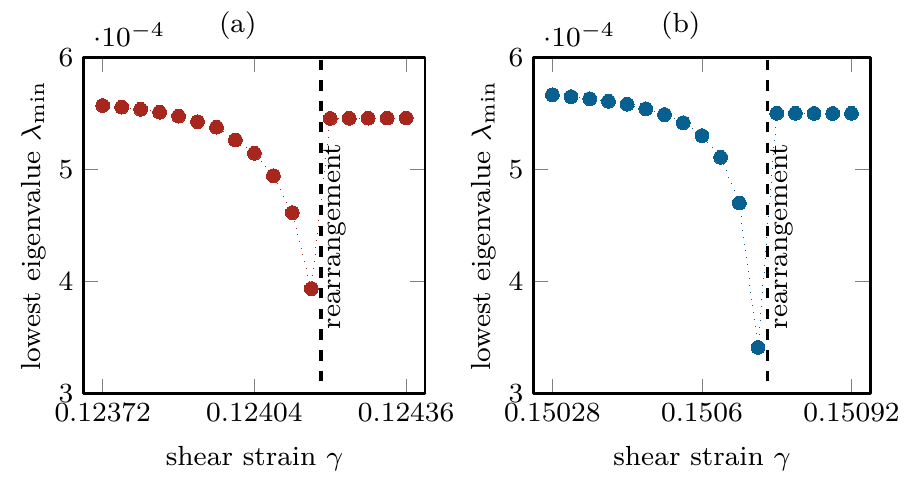}
\caption{Lowest positive eigenvalue $\lambda_{\text{min}}$ of the Hessian matrix $\boldsymbol{H}$ as a function of the shear strain $\gamma$ for the elementary plastic event E$_1$ in panel (a), and for the event E$_2$ in panel (b). For both events, $\lambda_{\text{min}}$ approaches zero through a square-root singularity.}
\label{fig:min_eig}
\end{figure}

To further study the two kinds of events observed in the model 2D network glass, we, next investigate the Hessian matrix $\boldsymbol{H}$ of the system. To this end, the eigenvalues $\lambda_i$ of the Hessian matrix $\boldsymbol{H}$ are calculated as functions of the shear strain $\gamma$. The elements of the Hessian matrix $H_{ij}$, given in Eq. \ref{eq:hessian}, are computed by using the finite-difference method. We follow the procedure presented in the supplemental material of Ref. \cite{Bonfanti2019}, applying a displacement increment of $10^{-7} \r{A}$. Figure \ref{fig:min_eig} depicts the evolution of the lowest positive eigenvalue $\lambda_{\text{min}}$ of the Hessian matrix $\boldsymbol{H}$ with increasing strain for the events E$_1$ and E$_2$. From this figure, it can clearly be seen, that for both events, the lowest eigenvalues $\lambda_{\text{min}}$ approaches zero via a square-root singularity, given in Eq. \ref{eq:lambda}. Thus, following the deformation theory of amorphous solids, both events are clearly identified as plastic instabilities under applied shear strain $\gamma$, occurring through a saddle-node bifurcation \cite{Rodney2011}. However, we are still curious about the fundamental difference between the two kinds of elementary plastic events.

\begin{figure*}[]
\centering
\includegraphics[width=17.2cm]{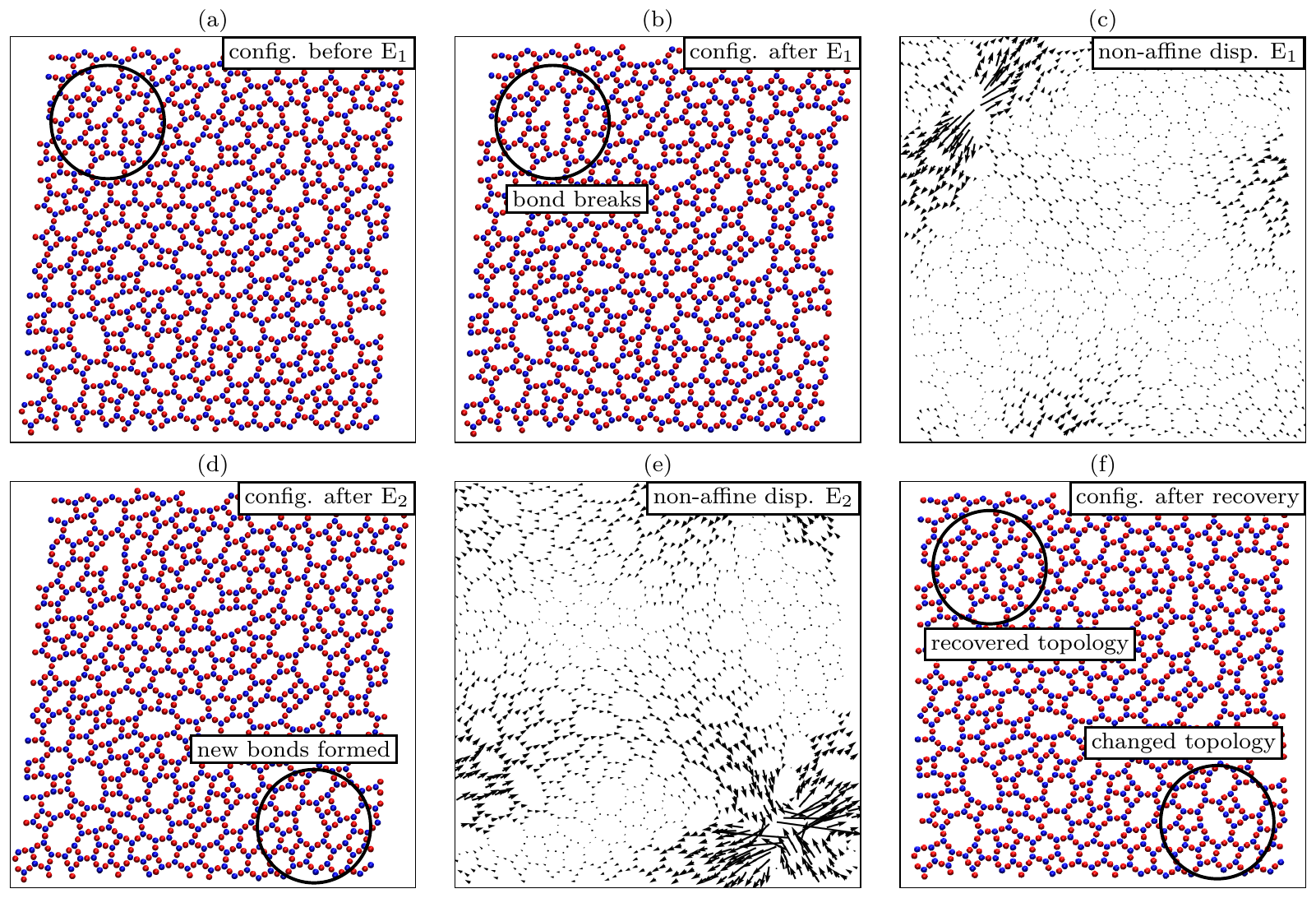}
\caption{The atomic configuration before the event E$_1$ is shown in (a), and after E$_1$ in panel panel (b). The nonaffine displacement field at E$_1$ is depicted in panel (c). The atomic configuration after E$_2$ is shown in panel (d), and the nonaffine displacement field at E$_2$ in panel (e). The atomic configuration after the recovery of E$_1$ is depicted in panel (f). The blue spheres represent silicon atoms, while the red spheres represent oxygen atoms. The encircled areas illustrate the local changes in the network structure due to plastic deformation. The size of the arrows in panels (c) and (e) is proportional to the magnitude of atomic displacement vectors.}
\label{fig:plastic_events}
\end{figure*}

The picture becomes clear by observing the network structure of the model 2D network glass before and right after the plastic instabilities. For this purpose, the atomic configuration before the event E$_1$ is depicted in Fig. \ref{fig:plastic_events}(a), and after E$_1$ in Fig. \ref{fig:plastic_events}(b). From these two figures it becomes clear that the first event E$_1$ corresponds to a localized rearrangement in the form of a bond-breaking event that results in the formation of a void and dangling bonds, while the bond orientation is barely influenced. The inspection of the atomic configuration after the event E$_2$, shown in Fig. \ref{fig:plastic_events}(d), reveals that this event corresponds to a rearrangement in the microstructure that changes the local network topology, i.e., bonding configuration. Unlike in the case of the event E$_1$, here, no voids or dangling bonds are observed after the plastic deformation. Instead, the bond orientations change to a significant degree. It is observed that the network of the model 2D glass is still fully coordinated, while the local ring neighborhood and size distribution is changed during this plastic event, as illustrated by the circled area in Fig. \ref{fig:plastic_events}(d).

These intriguing observations explain why the first event E$_1$ is recovered upon unloading, while the second event E$_2$ is not. In case of the event E$_1$, the broken bond retains its orientation when the void is formed. However, it can be reformed during shear unloading, such that the initial local ring structure is recovered, as can bee seen from Fig. \ref{fig:plastic_events}(f). In case of event E$_2$, we observe bonds that are broken and new bonds that are formed in the perpendicular direction. The local ring transformation is irreversible and cannot recover upon unloading. The nonaffine displacement fields belonging to E$_1$, shown in Fig. \ref{fig:plastic_events}(c), and to E$_2$, shown in Fig. \ref{fig:plastic_events}(e), indicate the localization of the eigenfunctions associated with the lowest eigenvalues $\lambda_{\text{min}}$ vanishing at the plastic instabilities.

\begin{figure}[]
\centering
\includegraphics[width=8.6cm]{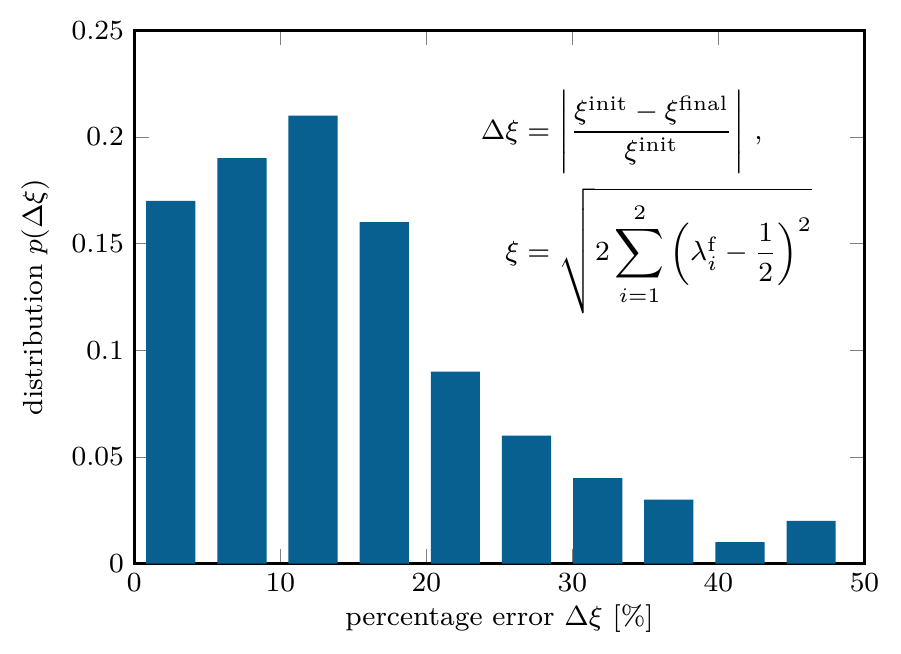}
\caption{Distribution of the percentage error of the orientation parameter $\Delta \xi$ for all identified  E$_2$-type, i.e., irreversible, events. The nonzero values clearly indicate localized changes in the glass network due to broken bonds in the one direction, while new bonds are formed in the perpendicular direction.}
\label{fig:hist}
\end{figure}

To underline the above findings, we perform an additional statistical analysis over the 100 samples. Each samples is subjected to simple shear loading and unloading until the first initial irreversible plastic event is triggered. Subsequently, for each identified event, we compare the network topology at the initial undeformed state with the final topology corresponding to the stress-free state of the blue unloading curve, shown in Fig. \ref{fig:sig_eps_sample_1_reload}. Rountree et al.~quantified the bond orientation in the network of bulk silica by using the fabric \mbox{tensor $\boldsymbol{F} = \langle \boldsymbol{n} \otimes \boldsymbol{n} \rangle$}, which measures the mean orientation of the unit normal vectors between silicon atoms $\boldsymbol{n}$  by averaging over their dyadic product \cite{Rountree2009}. Following this line of thought, we define an orientation parameter $\xi$ depending on the eigenvalues of the fabric tensor $\lambda^{\text{f}}_i$:
\begin{equation}
    \xi = \sqrt{2 \sum_{i=1}^{2} \left( \lambda^{\text{f}}_i-\frac{1}{2} \right)^2} \,.
\end{equation}
In case of isotropy it holds $\xi=0$, i.e., $\lambda^{\text{f}}_1 = \lambda^{\text{f}}_2 = \frac{1}{2}$. In case of anisotropy it holds $\xi=1$, i.e., $\lambda^{\text{f}}_1 = 1, \, \lambda^{\text{f}}_2 = 0$. The initial and final network topologies are compared by measuring the percentage error $\Delta \xi = \left| \frac{ \xi^{\text{init}} - \xi^{\text{final}} }{\xi^{\text{init}}} \right|$. To exclusively capture the nonaffine contribution, $\Delta \xi$ is calculated in a circular area with radius of $8$ \r A around the rearrangement center. We observe that for all reversible E$_1$-type events, the initial network topology is recovered during shear unloading, i.e., the initial bonds represented by the vectors $\boldsymbol{n}$ are rebuilt.  Approximately 46 \% of all events are reversible, thus, observed with $\Delta \xi = 0$. Based on all remaining E$_2$-type, i.e., irreversible, events, the probability distribution of the percentage error of the orientation \mbox{parameter $p(\Delta \xi)$}  is determined and presented in \mbox{Fig. \ref{fig:hist}}.  For each irreversible event, $\Delta \xi$ is found to be nonzero, where the range of values is approximately between 0 \% and 50 \%. The distribution $p(\Delta \xi)$ has a peak at a $\Delta \xi$ value of 12 \% and decreases strongly with respect to very large values of $\Delta \xi$. Since, for all E$_2$-type events $\Delta \xi$ has nonzero values, we hereby clearly identify localized changes in the glass network due to the irreversible events. In other words, the initial set of vectors $\boldsymbol{n}$, measured inside the local circular area, has changed. Thus, the statistical results underline that the event E$_2$, representatively depicted in Fig. \ref{fig:sig_eps_sample_1_reload}, is not just an exception but a typical case of an irreversible event that is accompanied by topological changes in the glass network. We elucidate that some bond-breaking events triggered under loading may also be irreversible. Upon shear unloading, the dangling atoms may bind to atoms other than the originally bonded atoms, leading to a locally altered network topology. Thus, irreversible events are always accompanied by the formation of new bonds due to mechanical loading.

The irreversible event E$_2$, observed here, can be directly linked to an experimental study on real 2D silica glass. Huang et al.~were able to excite and image atomic rearrangements in 2D silica glass, revealing that plastic deformation is accompanied by local ring transformations \cite{Huang2013}. A recent numerical study has reported two elementary plastic events in bulk silica glass undergoing AQS shear deformation, one event associated with a localized rearrangement without any change in the atomic coordination, the other resulting from an observed bond breaking \cite{Bonfanti2019}. However, in this study, we show that reversing the simple shear deformation is indispensable to reveal the fundamental mechanisms that underlie elementary plastic events in network glasses. Other studies have also reported that the directional covalent bonds between atoms are of particular importance in the plasticity of network glasses \cite{Rountree2009,Koziatek2015,Bamer2019a}. Our results confirm that the network structure plays a key role in understanding the fundamental mechanism of elementary plastic events in covalent glasses.

It should be noted that using the force-shifted potential function, given in Eq. (\ref{eq:force-shifted}), avoids only discontinuous energy and force at $r_{ij}=r_c$. However, a discontinuity appears in the Hessian matrix, since the latter is the second-order derivative of the potential energy. To avoid this, several methods have been proposed in the literature, e.g., Refs. \cite{Dubey2015,Chattoraj2019}. Thus, regarding future studies, it would be interesting to investigate whether the discontinuity of the Hessian at the cut-off distance affects the results in any way.

\section{CONCLUSIONS}

In summary, we have investigated the origin of the plastic response of a model 2D network glass subjected to athermal quasistatic simple shear deformation. Shear loading of the network glass revealed several stress drops, some that can be recovered and some not, when the shear deformation is reversed right after the stress drops. The events have been identified as the elementary building blocks in glassy solids, showing that, for each event, the lowest eigenvalue of the Hessian matrix approaches zero via a square-root singularity. Careful inspection of the atomic configurations before and after the plastic events revealed their underlying topological origin. In case of the reversible event, a bond is broken, while retaining its orientation and forming a void in the network. During shear unloading, the broken bond has been shown to reform, such that the initial local ring structure is recovered. In case of the irreversible event, local ring transformations have been observed, while the network remains fully coordinated. Bonds break and new bonds are formed in the perpendicular direction, i.e., the local bond orientations change to a significant degree. This event cannot recover upon unloading. We have shown that the two-dimensional framework of a network glass, composed of rings and ring neighborhoods, constitutes a promising tool that can provide further insights into the elementary mechanisms of plastic deformation in disordered solids.




\begin{thebibliography}{10}

\bibitem{Spaepen1977}
F.~Spaepen, Acta Metall. 25 (1977) 407.

\bibitem{Argon1979b}
A.~Argon, Acta Metall. 27 (1979) 47.

\bibitem{Falk1998}
M.L.~Falk, J.S.~Langer, Phys. Rev. E 57 (1998) 7192.
  
\bibitem{Tomida1993}
T.~Tomida, T.~Egami, Phys. Rev. B 48 (1993) 3048.

\bibitem{Lacks2004}
D.J.~Lacks, M.J.~Osborne, Phys. Rev. Lett. 93 (2004) 255501.

\bibitem{Maloney2006} 
C.E.~Maloney, A.~Lema{\^i}tre, Phys. Rev. E 74 (2006) 016118.

\bibitem{Tanguy2006}
A.~Tanguy, F.~Leonforte, J.L.~Barrat, Eur. Phys. J. E: Soft Matter Biol. Phys. 20 (2006) 355. 
 
\bibitem{Tsamados2009}
M.~Tsamados, A.~Tanguy, C.~Goldenberg, J.L.~Barrat, Phys. Rev. E 80 (2009) 026112.

\bibitem{Patinet2016}
S.~Patinet, D.~Vandembroucq, M.L.~Falk, Phys. Rev. Lett. 117 (2016) 045501.

\bibitem{Barbot2018}
A.~Barbot, M.~Lerbinger, A.~Hernandez-Garcia, R.~Garc{\`i}a-Garc{\`i}a, M.L.~Falk, D.~Vandembroucq, S.~Patinet,  Phys. Rev. E 97 (2018) 033001.

\bibitem{Heuer2008}
A.~Heuer, J Phys: Condens. Matter. 20 (2008) 373101.

\bibitem{Rodney2011}
D.~Rodney, A.~Tanguy, D.~Vandembroucq, Model. Simul. Mater. Sci. Eng. 19 (2011) 083001.

\bibitem{Hufnagel2016}
T.~Hufnagel, C.~Schuh, M.L.~Falk, Acta Mater. 109 (2016) 375.

\bibitem{Ashwin2013}
A.~Joy, O.~Gendelman, I.~Procaccia, C.~Shor, Phys. Rev. E 88 (2013) 022310.

\bibitem{Gendelman2015}
O.~Gendelman, P.~Jaiswal, I.~Procaccia, B.~Gupta, J.~Zylberg, Europhys. Lett. 109 (2015) 16002.

\bibitem{Eshelby1957}
J.~Eshelby, Proc. R. Soc. London, Ser. A 241 (1957) 376.

\bibitem{Albaret2016}
T.~Albaret, A.~Tanguy, F.~Boioli, D.~Rodney, Phys. Rev. E 93 (2016) 053002.

\bibitem{Hieronymus_Schmidt2017}
V.~Hieronymus-Schmidt, H.~R{\"o}sner, G.~Wilde, A.~Zaccone, Phys. Rev. B 95 (2017) 134111.

\bibitem{Cao2018}
X.~Cao, A.~Nicolas, D.~Trimcev, A.~Rosso, Soft Mater 14 (2018) 3640.

\bibitem{Maloney2004a}
C.~Maloney, A.~Lema{\^i}tre, Phys. Rev. Lett. 93 (2004) 195501.

\bibitem{Maloney2004b}
C.~Maloney, A.~Lema{\^i}tre, Phys. Rev. Lett. 93 (2004) 016001.

\bibitem{Lemaitre2007}
A.~Lemaitre, C.~Caroli, Phys. Rev. E 76 (2007) 036104.

\bibitem{Xu2017}
B.~Xu, M.L.~Falk, J.~Li, L.~Kong, Phys. Rev. B 95 (2017) 144201.

\bibitem{Xu2018}
B.~Xu, M.L.~Falk, J.F.~Li, L.T.~Kong, Phys. Rev. Lett. 120 (2018) 125503.

\bibitem{Argon1983}
A.~Argon, L.~Shi, Acta. Metall. 31 (1983) 499.
  
\bibitem{Morley2018}
D.~Morley, M.~Wilson, J. Phys.: Condens. Matter 30 (2018) 50LT02.

\bibitem{Morley2019}
D.~Morley, M.~Wilson, Mol. Phys. 117 (2019) 3148.

\bibitem{Bamer2020}
F.~Bamer, F.~Ebrahem, B.~Markert, Materialia 9 (2020) 100556.

\bibitem{Lichtenstein2012a}
L.~Lichtenstein, C.~B{\"u}chner, B.~Yang, S.~Shaikhutdinov, M.~Heyde, M.~Sierka, R.~W{\l}odarczyk, J.~Sauer, H.J.~Freund, Angew. Chem., Int. Ed. 51 (2012) 404.

\bibitem{Roy2018}
P.~Roy, M.~Heyde, A.~Heuer, Phys. Chem. Chem. Phys. 20 (2018) 14725--14739.

\bibitem{Roy2019a}
P.K.~Roy, A.~Heuer, Phys. Rev. Lett. 122 (2019) 016104.

\bibitem{Roy2019b}
P.~Roy, A.~Heuer, J. Phys.: Condens. Matter 31 (2019) 225703.

\bibitem{Bamer2019b}
F.~Bamer, F.~Ebrahem, B.~Markert, Comp. Mater. Sci. 163 (2019) 301.

\bibitem{Ebrahem2020}
F.~Ebrahem, F.~Bamer, B.~Markert, J. Mater. Sci. 55 (2020) 3470.

\bibitem{Zachariasen1932}
W.~Zachariasen, J. Am. Chem. Soc. 54 (1932) 3841.

\bibitem{Allen1989}
M.~Allen, D.~Tildeslay, \textit{Computer Simulation of Liquids} (Clarendon Press, London, 1989).

\bibitem{Todorova2006}
T.K.~Todorova, M.~Sierka, J.~Sauer, S.~Kaya, J.~Weissenrieder, J.L.~Lu, H.J.~Gao, S.~Shaikhutdinov, H.J.~Freund, Phys. Rev. B 73 (2006) 165414.

\bibitem{Ebrahem2020b}
F.~Ebrahem, F.~Bamer, B.~Markert, Mat. Sci. Eng. A 780 (2020) 139189.

\bibitem{Aboav1970}
DA.~Aboav, Metallography 3 (1970) 383.

\bibitem{Aboav1980}
DA.~Aboav, Metallography 13 (1980) 43.

\bibitem{Weaire1974}
D.~Weaire, Metallography 7 (1974) 157.

\bibitem{Talamali2011}
M.~Talamali, V.~Pet{\"a}j{\"a}, D.~Vandembroucq, S.~Roux, Phys. Rev. E 84 (2011) 016115.

\bibitem{Nicolas2018}
A.~Nicolas, E.E.~Ferrero, K.~Martens, J.L.~Barrat, Rev. Mod. Phys. 90 (2018) 045006.

\bibitem{Rountree2009}
C.L.~Rountree, D.~Vandembroucq, M.~Talamali, E.~Bouchaud, S.~Roux, Phys. Rev. Lett. 102 (2009) 195501.

\bibitem{Bonfanti2019}
S.~Bonfanti, R.~Guerra, C.~Mondal, I.~Procaccia, S.~Zapperi, Phys. Rev. E 100 (2019) 060602(R).

\bibitem{Huang2013}
P.Y.~Huang, S.~Kurasch, J.S.~Alden, A.~Shekhawat, A.A.~Alemi, P.L.~McEuen, J.P.~Sethna, U.~Kaiser, D.A.~Muller, Science 342 (2013) 224.


\bibitem{Koziatek2015}
P.~Koziatek and J.L.~Barrat, D.~Rodney, J. Non-Crystal. Solids 414 (2015) 7. 

\bibitem{Bamer2019a}
F.~Bamer, F.~Ebrahem, B.~Markert, J. Non-Crystal. Solids 503 (2019) 176.

\bibitem{Dubey2015}
A.K.~Dubey, H.G.E.~Hentschel, P.K.~Jaiswal, C.~Mondal, I.~Procaccia, B.S.~Gupta, Europhys. Lett. 112 (2015) 17011.

\bibitem{Chattoraj2019}
J.~Chattoraj, O.~Gendelman, M.P.~Ciamarra, I.~Procaccia, Phys. Rev. E 100 (2019) 042901.
\end{thebibliography}
\end{document}